\documentclass{article}

\usepackage{arxiv}

\usepackage[utf8]{inputenc} % allow utf-8 input
\usepackage[T1]{fontenc}    % use 8-bit T1 fonts
\usepackage{hyperref}       % hyperlinks
\usepackage{url}            % simple URL typesetting
\usepackage{booktabs}       % professional-quality tables
\usepackage{amsfonts}       % blackboard math symbols
\usepackage{nicefrac}       % compact symbols for 1/2, etc.
\usepackage{microtype}      % microtypography
\usepackage{lipsum}		% Can be removed after putting your text content
\usepackage{graphicx}
\usepackage{doi}

\title{Comparing Software Developers with ChatGPT: \\An Empirical Investigation}

\author{ \href{https://orcid.org/0000-0002-4388-6572}{\includegraphics[scale=0.06]{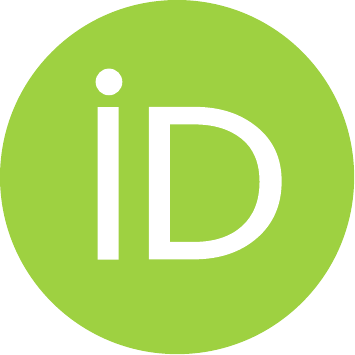}\hspace{1mm}Nathalia Nascimento} \\
	David R. Cheriton School of Computer Science, \\
	University of Waterloo, (UW)\\
	Waterloo, N2L 3G1, Canada \\
	\texttt{nmoraesd@uwaterloo.ca} \\
	\And
	{Paulo Alencar} \\
	David R. Cheriton School of Computer Science, \\
	University of Waterloo, (UW)\\
	Waterloo, N2L 3G1, Canada \\
	\texttt{palencar@uwaterloo.ca} \\
 \And
	{Donald Cowan} \\
	David R. Cheriton School of Computer Science, \\
	University of Waterloo, (UW)\\
	Waterloo, N2L 3G1, Canada \\
	\texttt{dcowan@uwaterloo.ca} \\
}

\date{}

\renewcommand{\shorttitle}{ChatGPT vs. Human Programmers}

\hypersetup{
pdftitle={ChatGPT vs Human Programmers},
pdfsubject={q-bio.NC, q-bio.QM},
pdfauthor={Nathalia Nascimento, Paulo Alencar, Donald Cowan},
pdfkeywords={Software Engineering; AI-based solutions; Performance Evaluation; ChatGPT; Machine Learning},
}

\begin{document}
\maketitle

\begin{abstract}
	The advent of automation in particular Software Engineering (SE) tasks has transitioned from theory to reality. Numerous scholarly articles have documented the successful application of Artificial Intelligence to address issues in areas such as project management, modeling, testing, and development. A recent innovation is the introduction of ChatGPT, an ML-infused chatbot, touted as a resource proficient in generating programming codes and formulating software testing strategies for developers and testers respectively. Although there is speculation that AI-based computation can increase productivity and even substitute software engineers in software development, there is currently
a lack of empirical evidence to verify this. Moreover, despite the primary focus on enhancing the accuracy of AI systems, non-functional requirements including energy efficiency, vulnerability, fairness (i.e., human bias), and safety frequently receive insufficient attention. This paper posits that a comprehensive comparison of software engineers and AI-based solutions, considering various evaluation criteria, is pivotal in fostering human-machine collaboration, enhancing the reliability of AI-based methods, and understanding task suitability for humans or AI. Furthermore, it facilitates the effective implementation of cooperative work structures and human-in-the-loop processes. This paper conducts an empirical investigation, contrasting the performance of software engineers and AI systems, like ChatGPT, across different evaluation metrics. The empirical study includes a case of assessing ChatGPT-generated code versus code produced by developers and uploaded in Leetcode. 
\end{abstract}

% keywords can be removed

\section{Introduction}

The popularity of AI-based tools such as ChatGPT (versions 3 and 4) \cite{openai2023gpt4,eloundou2023gpts}, a tool made available by OpenAI, has exploded. ChatGPT has set a record for the fastest-growing user, having reached 100 million users in January 2023 with 25 million daily visitors. As a result, the recent automation capabilities supported by ChatGPT resulted in increased interest in topics such as the potentially increasing the automation of software development tasks such as coding and testing \cite{white2023chatgpt,sobania2023analysis,sarro2023automated} 
enabling programmers to make their tasks more efficient\cite{eloundou2023gpts} and allowing them to be more productive, and assessing how human-machine teams that can be more effective in software development tasks \cite{melo2023supporting,imai2022github}. Indeed, tools such as ChatGPT have indeed led to impressive results both in terms of quantity and quality, and the produced outcomes (e.g., code) are in many cases comparable to what is produced by humans. For example, Golzadeh et al. performed an investigation in large open-source projects on GitHub and observed that bots are among the most active contributors, without being labeled as bots \cite{golzadeh2022recognizing}. 
%This raises questions about how  software developers and machine-learning algorithms can be compared when executing the same task. 

However, while the research community has put in a great deal of effort to enhance the accuracy of AI-based approaches, they have often overlooked other non-functional requirements \cite{bender2021dangers}, such as energy efficiency \cite{georgiou2022green}, vulnerability \cite{pearce2022asleep}, fairness (i.e. human bias) \cite{10.1145/3457607}, and safety \cite{sarro2023automated,chen2021evaluating}. According to Georgiou et al.\cite{georgiou2022green}, the use of deep learning frameworks has considerably increased energy consumption and CO2 emissions. For example, only ``ChatGPT has been estimated to consume the equivalent of the electricity consumed by 175,000 people in Denmark per month" \cite{sarro2023automated}. Pearce et al. \cite{pearce2022asleep} investigated the use of an AI-based code generator in security-relevant scenarios, and observed that 40\% of the provided solutions were vulnerable. Sarro \cite{sarro2023automated} brings attention to the presence of bias in various real-world systems relying on ML, such as advertisement and recruitment processes, and to safety problems, such as insecure code generation \cite{chen2021evaluating}, as creating novel malware or inserting malware into the generated system, performance of dangerous operation such as file manipulation \cite{inala2022fault}, or facilitating harmful acts, such as scamming, harassment, misinformation, and election interference. According to Sarro \cite{sarro2023automated}, not even a software engineer, regardless of their level of expertise, would be able to manually detect all possibilities for improving these non-functional characteristics.

In addition, while there is speculation that AI-based computation can increase productivity and even substitute software engineers in software development, there is currently a lack of empirical evidence to verify this \cite{imai2022github}. In fact, there are few papers providing empirical studies to investigate the use of machine-learning techniques in Software Engineering and compare an ML-based solution against those provided by software engineers, particularly considering different non-functional properties. For example, Nascimento et al. \cite{nascimento2018toward} present an empirical study to compare software engineers to machine learning in the domain of the Internet of Things (IoT), addressing performance and reuse criteria, and conclude that ``we cannot state that ML improves the performance of an application in comparison to solutions provided by IoT expert software engineers... Our experiment indicates that in some cases, software engineers outperform machine-learning algorithms, whereas in other cases, they do not."

Such an understanding is essential in realizing novel human-in-the-loop approaches in which AI procedures assist software developers in achieving tasks. Human-in-the-loop approaches, which take into account the strengths and weaknesses of humans and AI solutions, are fundamental not only to providing a basis for cooperative human-machine work or teams not only in software engineering but also in other application areas.

This paper presents an empirical study  \cite{easterbrook2008selecting} to compare how software engineers and AI systems can be compared with respect to non-functional requirements such as performance and memory efficiency. The empirical study involves a case study assessing ChatGPT-generated code versus code produced by developers and uploaded in Leetcode, which consists of three steps: (i) we selected a contest from Leetcode that contains programming problems at different difficulty levels; (ii) we used these problems as prompts to ChatGPT to generate code; and (iii) we uploaded the ChatGPT code solution to Leetcode and compared them to the previous solutions based on performance and efficiency metrics.

This paper is organized as follows. Section 2 presents the related work. Section 3 presents the empirical study, describing research questions, hypotheses, and the objective of the study. Section 4 presents the experimental results and threats to validity. The paper ends in Section 5 with concluding remarks and suggestions for future work.

\section{Related Work} \label{sec:relatedwork}

 Imai \cite{imai2022github} claims that while there is speculation that AI-based computation can increase productivity and even substitute human pair programmers in software development, there is currently a lack of empirical evidence to verify this. In fact, there are few papers providing empirical studies to investigate the use of machine-learning techniques in Software Engineering and compare an ML-based solution against those provided by software engineers, particularly considering non-functional properties \cite{nascimento2018toward}. Imai \cite{imai2022github} conducted an empirical study to compare the productivity and code quality between pair programming with GitHub Copilot and human pair programming. GitHub Copilot, a tool launched by OpenAI and GitHub, to provide code snippets and automatically fill in parts of code, gives users the choice to accept or reject its assistance depending on their knowledge. The experiment involved 21 participants, with each one receiving a project to code and a developer partner, either human or the GitHub Copilot. To evaluate, Imai \cite{imai2022github} calculated productivity based on the number of lines of code added and code quality based on the number of lines of code removed after being added. The results showed that programming with Copilot helps generate more lines of code than human pair-programming in the same period of time, but the code quality was lower. Additionally, the author performed a preliminary evaluation of code confiability, as they reported that programmers tend to inspect the code generated by AI less than human pair-programmers.

Nguyen and Nadi \cite{nguyen2022empirical} also conducted an empirical study using GitHub Copilot's generated code to assess the correctness and understandability of solutions for 33 Leetcode problems in four different programming languages. To evaluate the correctness, the authors counted the number of test cases that passed for each problem, and to assess understandability, they employed SonarQube, an open-source platform for static code analysis, to calculate complexity and cognitive complexity metrics. The authors did not focus on performance and memory efficiency, so they did not provide execution time or memory use for each solution, nor did they compare the Copilot's and human-written solutions.

 Li et al. presented AlphaCode \cite{li2022competition}, a code generation system. They trained their model using GitHub and CodeContests data. After using AlphaCode to solve competitive programming problems from the Codeforces platform, the authors state that ``AlphaCode achieved on average a ranking of top 54.3\% in competitions with more than 5,000 participants". They compared their solution against the developers' ones based on the contest metrics, which are a fraction of the time left in the contest and incorrect submission penalties.
%number of submission trials for each question. 
%They also performed an empirical study to investigate AI versus human coding.
 Lertbanjongngam et al. \cite{lertbanjongngam2022empirical} extended AlphaCode evaluation. Using the same code provided by AlphaCode in \cite{li2022competition} for the Codeforces competitive programming problems, they assessed human-like coding similarity, performance, and memory efficiency. Their results show that AlphaCode-generated codes are similar to human codes, having a uniqueness of less than 10\% of code fragments. They also show that the code that was produced exhibits similar or inferior execution time and memory usage compared to the code written by humans. In contrast to Lertbanjongngam et al. \cite{lertbanjongngam2022empirical}, we utilized ChatGpt and randomly selected coding problems for our first experiment, where we compared the performance of code generators to that of human solutions. The tool was employed to generate code for each selected problem. 

 Nascimento et al. \cite{nascimento2018toward} conducted an empirical study to investigate the automation of a coding problem without a set of unit tests. Their experiment does not rely on match-based metrics such as generating a solution that passes a specific set of unit tests; instead, it is based on functional correctness, emphasizing that the solution should work efficiently. In addition, their experiment employs an unsupervised ML-based approach, where the ML approach offers a solution to the problem before it is presented to software engineers.

\section{Experiment: ChatGPT vs Programmers - An Empirical Study Addressing Performance and Efficiency}

Software engineers are constantly evaluated by their capability of problem-solving coding, which involves creating a program to solve a problem based on the problem specifications. Given a problem, they need to be able to understand what is being asked and write code to solve the problem. For years, this kind of task has been used to rank software engineers during job interviews and programming contests. Their evaluation usually relies on the number of right solutions, the performance of the solutions, memory efficiency, and development time.

For almost 76 years, researchers have been writing about the concept of ``automatic programming" \cite{brooks1987no}. Recent advancements in language models, such as the Generative Pre-trained Transformer (GPT) series, have greatly advanced the field of automatic programming by enabling the generation of code, program synthesis, code completion, and bug detection. GPTs are deep neural networks that are pre-trained on vast amounts of natural language text and then fine-tuned for specific tasks such as question answering. Given a natural language description of a task, GPTs can generate a code that accomplishes the desired task, which is achieved by fine-tuning the model on a large corpus of code.

OpenAI's Codex is a language model built on the GPT architecture and integrated into ChatGPT. Chen et al. \cite{chen2021evaluating} introduced Codex and used a dataset of 163 coding problems to evaluate it, including introductory, interview, and competition problems. Their evaluation consisted of generating many solutions and checking if at least one passed the unit tests. Accordingly, after generating 100 samples per problem, their solution was able to solve 77.5\% of the problems. They evaluated their approach based on efficacy (number of tests passed), not accessing performance, and memory efficiency. Imai \cite{imai2022github}, and Nguyen and Nadi \cite{nguyen2022empirical} also performed empirical studies with a version of Codex. The non-functional requirements they evaluated are productivity, code quality, confiability, correctness, and understandability. We describe their findings in Section \ref{sec:relatedwork}.

\subsection{Objective}
\begin{figure*}[htb!]
	\centering
	\includegraphics[scale=0.40]{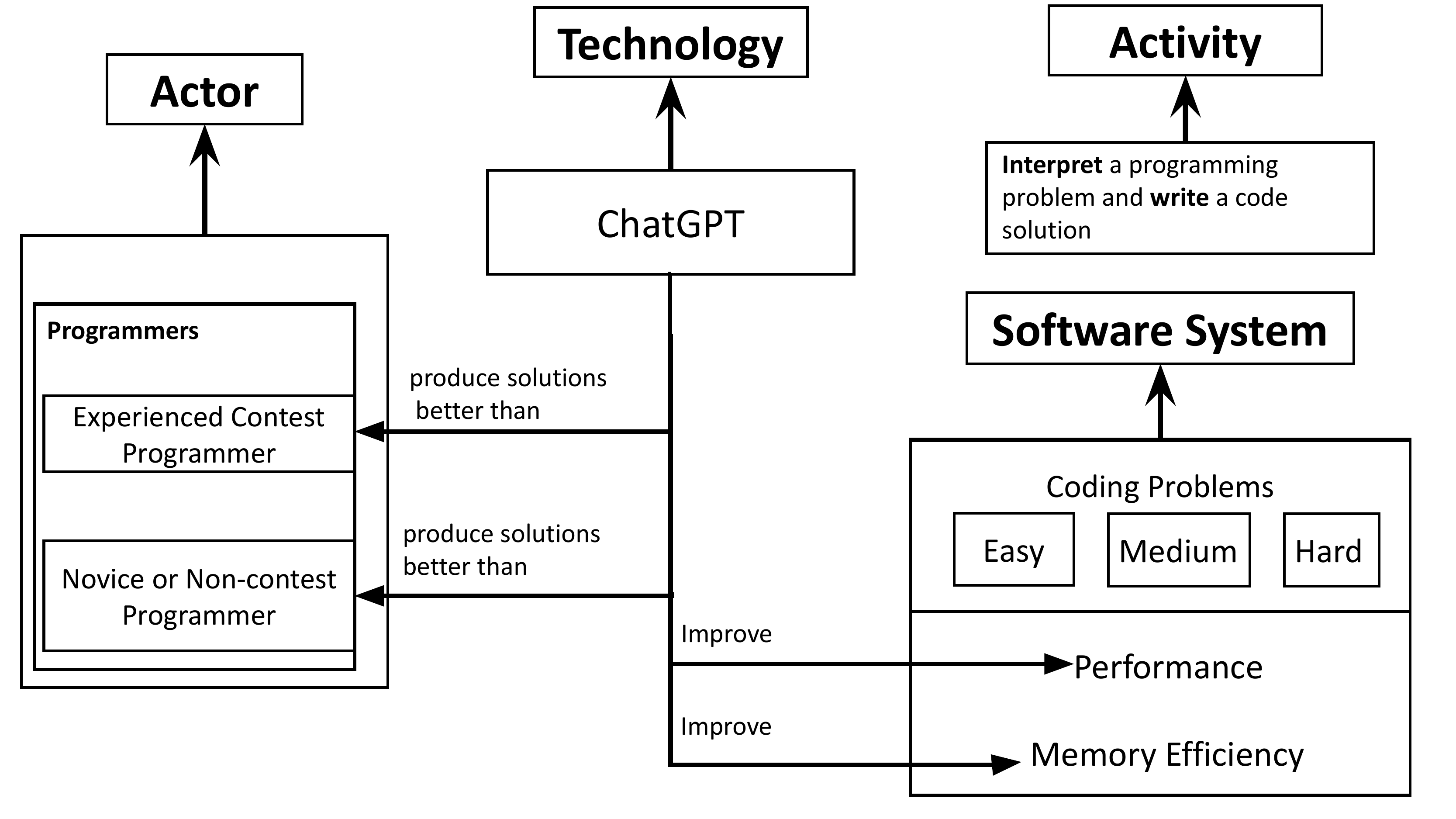} 
	\centering
	\caption{Theory \cite{sjoberg2008building}: ChatGPT outperforms programmers in problem-solving coding tasks with higher performance and memory efficiency.}
	\label{fig:theory1}
\end{figure*}

%“How does the problem-solving capacity of programmers compare to that of ChatGPT in terms of performance and memory efficiency, with regards to coding tasks?"
In this context, we decided to ask the following question: ``How do software engineers compare with AI solutions with respect to performance and memory efficiency?" To explore this question, we selected coding problems as our target activity and compared a solution provided by an experienced contest programmer with a solution provided by ChatGPT. In short, Figure \ref{fig:theory1} depicts the theory \cite{sjoberg2008building} that we investigate in this paper. According to the theory, the variables that we intend to isolate and measure are the performance and memory efficiency achieved from three kinds of solutions: i) solutions provided by ChatGPT; ii) solutions provided by experience contest programmers; and iii) solutions provided by novice or non-contest programmers.

To evaluate the relationship among these variables, we performed a controlled experiment using LeetCode, a well-established online platform for programming interview preparation and subsequently juxtaposed the solution presented by ChatGPT with those previously developed by software engineers. The next subsection describes the research questions (RQx) and the theory's propositions. 
To perform this experiment, we selected one of the most recent contests in LeetCode with novel coding problems \footnote{The contest description is available at:\\
	\href{https://leetcode.com/contest/biweekly-contest-103/}{https://leetcode.com/contest/biweekly-contest-103/ (accessed on May 15 2023)}}.
Then, we used these problems as prompts to ChatGPT to generate code. Subsequently, the code produced by ChatGPT was uploaded to Leetcode, where it was evaluated against pre-existing solutions, utilizing performance and efficiency metrics. Given that Leetcode compares the submitted solution with all prior solutions to the same problem, we initiated a discussion on these metrics, predicated on programmer expertise. To this end, we selected a group of 42 programmers, differentiated by their rank in the contest, and categorized them according to the number of programming contests they have already attended in Leetcode.

\subsection{Questions and Hypotheses} \label{sec:questionsexp1}

In the domain of problem-solving coding, how does the result from a ChatGPT solution differ from solutions provided by...

%while solving coding problems at
\begin{quote}
	{\bf RQ1}. ...{\bf experienced contest programmers} while solving easy, medium, or hard coding problems with respect to their {\bf performance}?
	
	{\bf RQ2}. ...{\bf experienced contest programmers} while solving easy, medium, or hard coding problems with respect to their {\bf memory efficiency}?
	
	{\bf RQ3}. ...{\bf novice contest programmers} while solving easy, medium, or hard coding problems with respect to their {\bf performance}?
	
	{\bf RQ4}. ...{\bf novice contest programmers} while solving easy, medium, or hard coding problems with respect to their {\bf memory efficiency}?
\end{quote}

Each RQ is based on one or more   hypotheses, which are described next.
{\bf H - RQ1}.
\begin{itemize}
	\item H0. ChatGPT does not improve the performance of coding solutions compared to solutions provided by experienced contest programmers.
	\item HA. ChatGPT improves the performance of coding solutions compared to solutions provided by experienced contest programmers.
\end{itemize}

{\bf H - RQ2}.
\begin{itemize}
	\item H0. ChatGPT does not improve the memory efficiency of coding solutions compared to solutions provided by experienced contest programmers.
	\item HA. ChatGPT improves the memory efficiency of coding solutions compared to solutions provided by experienced contest programmers.
\end{itemize}

{\bf H - RQ3}.
\begin{itemize}
	\item H0. ChatGPT does not improve the performance of coding solutions compared to solutions provided by novice contest programmers.
	\item HA. ChatGPT improves the performance of coding solutions compared to solutions provided by novice contest programmers.
\end{itemize}

{\bf H - RQ4}.
\begin{itemize}
	\item H0. ChatGPT does not improve the memory efficiency of coding solutions compared to solutions provided by novice contest programmers.
	\item HA. ChatGPT improves the memory efficiency of coding solutions compared to solutions provided by novice contest programmers.
\end{itemize}

\subsection{The Object of the Study: ChatGPT}

\begin{figure*}[htb!]
	\centering
	\includegraphics[scale=0.40]{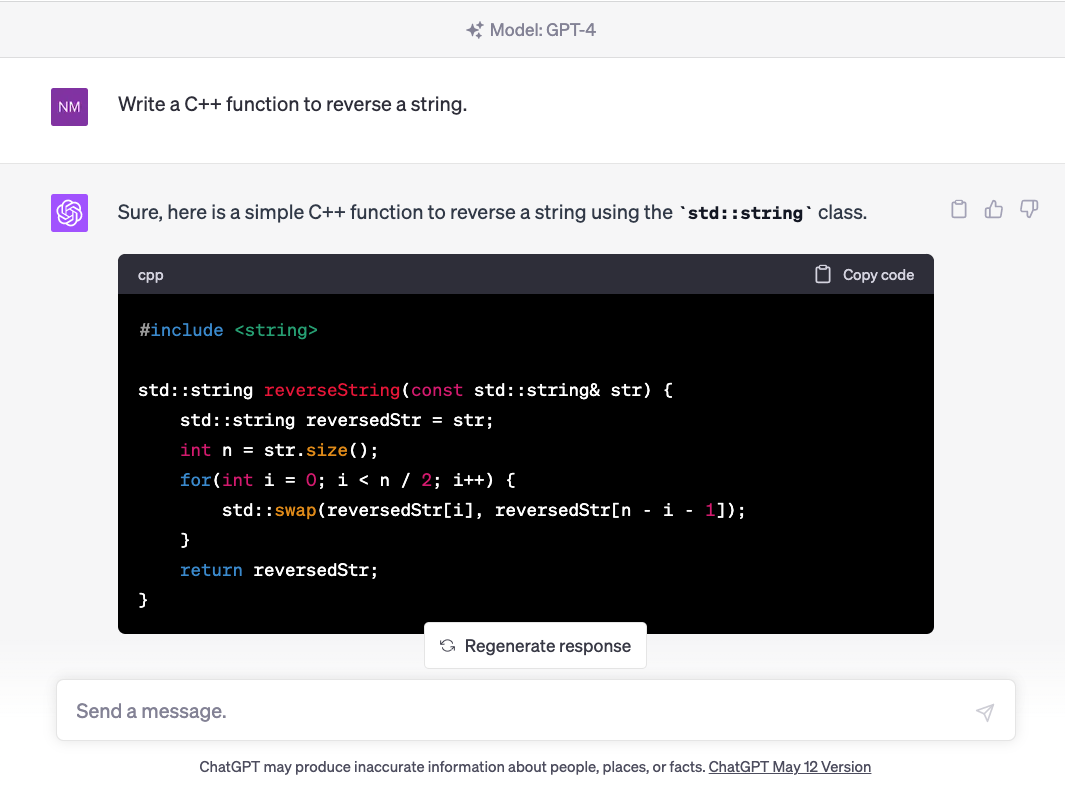} 
	\centering
	\caption{Example of using ChatGPT to solve a coding problem.}
	\label{fig:chatgpt}
\end{figure*}

By using ChatGPT, developers can generate code from natural language input, as shown in Figure \ref{fig:chatgpt}. In addition to the code, the system can also provide an explanation and evaluation for the generated code, such as analyzing time and space complexity. 

Although ChatGPT is capable of generating functional code for an expansive array of programming languages (including Java, Kotlin, Python, C++, JavaScript, TypeScript, PHP, Go-lang, Ruby, Swift, and more), we opted to focus on C++. This decision was taken by the fact that C++ is predominantly used by the most seasoned contest programmers. We used ChatGPT-4 to generate the code solutions. 
%Despite ChatGPT can generate functional code for a wide range of programming languages (Java, Kotlin, Python, C++, JavaScript, TypeScript, PHP, Go-lang, Ruby, Swift and more), Nguyen and Nadi performed an empirical experiment using Codex and observed that the system presents better results for Java development \cite{10.1145/3524842.3528470}. Thus, we will use Java in this experiment. 
%We find that Copilot’s Java suggestions have the highest correctness score (57%)

\subsection{Controlled Experiment} \label{sec:experiment}

The initial phase of the experiment involved choosing a platform to select and implement the coding solutions. Our objective was to present both sets of solutions to the same application and assess the outcomes using an identical evaluation process. We opted for Leetcode, an online platform for preparing for coding interviews. It has more than 4 million users and provides more than 2000 coding problems and solutions, usually used by big tech companies to assess developers' skills and conduct interviews. Leetcode compares solutions based on performance (runtime execution) and memory usage. 

We chose one of the most recent programming contests on Leetcode, which presented four unique programming challenges classified into one easy, two medium, and one hard problem. Out of the 17,137 participants, 12,493 managed to solve at least one problem. The number decreased to 10,733 for those who solved two problems, further reduced to 7,133 for three problems, and dwindled to a mere 700 for all four problems. 

As illustrated in Figure \ref{fig:theory1}, one of the factors we aim to isolate and quantify is the performance and memory efficiency of the solutions in relation to the programmer's skill level. To this end, we chose programmers based on their ranking in the contest, selecting those who used C++ and managed to solve at least three questions, ranging from the highest to the lowest rank. Utilizing LeetCode's parameters, we divided them into two categories: experienced contest programmers and novice or non-contest programmers. Ultimately, we juxtaposed the performance and memory efficiency of the solutions generated by ChatGPT against those provided by both groups of participants.

%Codes with more details and explanations use more memory. Systems that require more explainability, details, and traceability ? We tried to {\color{red} COMPLETE}
\subsubsection{Participant Analysis}

\begin{figure}[htb!]
	\centering
	\includegraphics[scale=0.45]{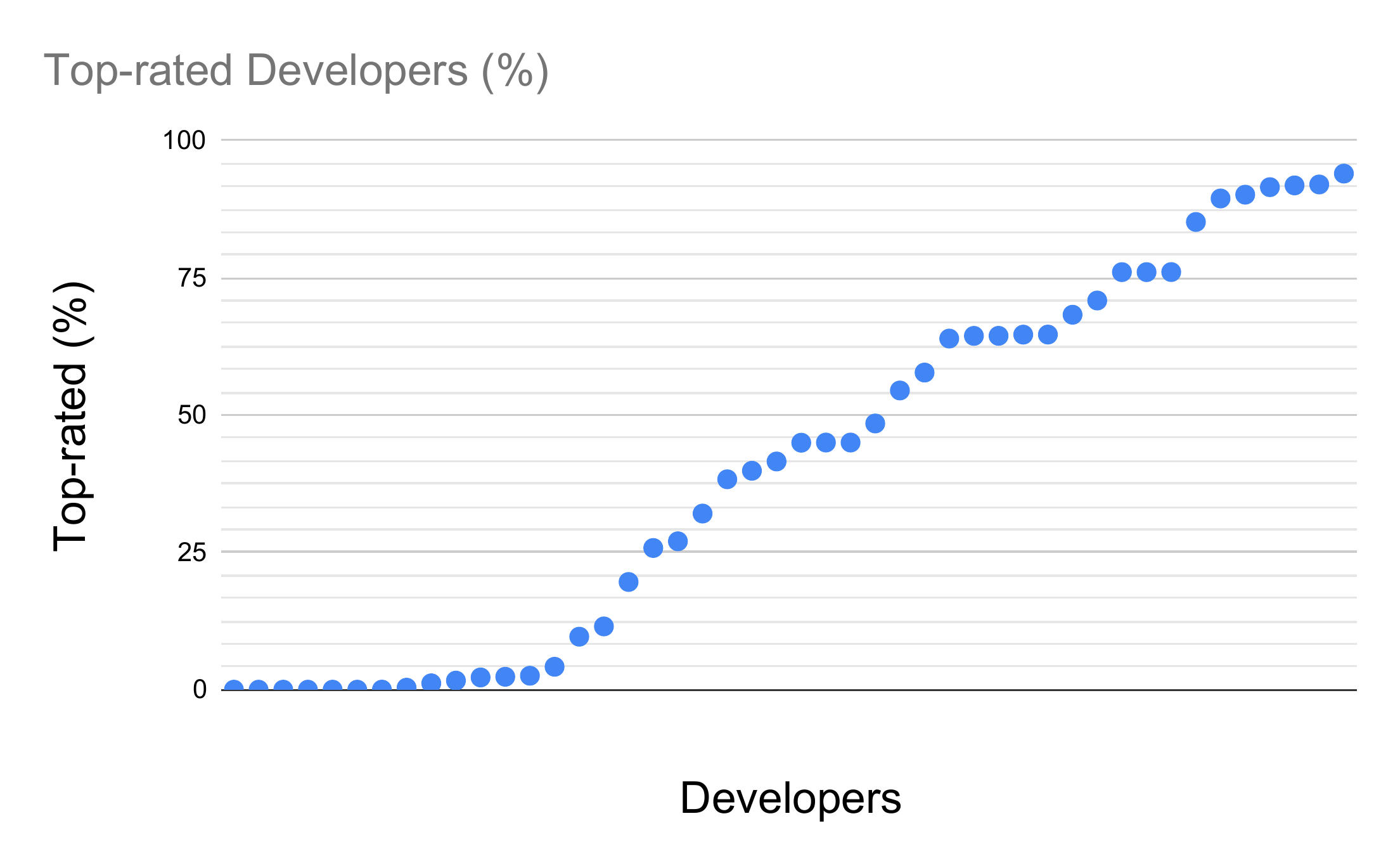} 
	\centering
	\caption{Distribution of participants based on LeetCode global ranking.}
	\label{fig:programmers}
\end{figure}

Figure \ref{fig:programmers} displays a distribution graph of the participants based on their ranking level on LeetCode, with lower levels indicating more experienced programmers. For instance, eight of the selected programmers fall within the top 1\% as per LeetCode's ranking.

Programmers in the top 30\% were grouped as Experienced Contest Programmers, skilled in problem-solving, algorithmic thinking, and coding languages from participating in numerous contests. They excel at quickly comprehending and solving problems under time pressure. The remaining participants were categorized as Novice or Non-contest Programmers, whose expertise lies more in software development tasks like design, coding, testing, and debugging. Despite their ability to solve complex problems, they may lack the speed and specialized algorithmic knowledge of contest programmers due to less exposure to competitive coding.

\subsubsection{Experiment: Solutions provided by ChatGPT}
We employed ChatGPT-4 to formulate a solution for each problem within the contest.\footnote{The prompt and code solutions provided are available at:\\
	\href{https://docs.google.com/spreadsheets/d/1Fqlz5kWQ7-Einxy5FRoLEeOXw_fBXulbO1qKbcorvyI/edit?usp=sharing}{Google Drive form (accessed on May 15 2023)}}. The original descriptions of the problems were used as prompts, without any textual modifications to simplify the problem interpretation.

Table \ref{tab:chatgptresults} encapsulates the list of problems along with the performance and memory efficiency of each solution. Upon running the solution on the LeetCode platform, it generates data on its runtime and memory efficiency. Furthermore, it provides a global ranking for each question, indicating the percentage of solutions outperformed by this solution in terms of performance and memory efficiency. These outcomes may fluctuate depending on LeetCode's processing demand. To ensure that the variation in time was inconsequential, we executed the same solution multiple times. We also evaluated the time and space complexity of the solutions. Interestingly, despite certain solutions demonstrating superior time complexity analysis, others showcased superior runtime performance and memory efficiency. This example is clearly demonstrated in the two solutions provided for the simplest problem. It is known that optimizing an application's performance depends not just on minimizing the algorithm's iterations, but also on selecting the most effective data structures.

\label{tab:chatgptresults}
\begin{table*}[!htb]

\scalebox{0.75}{% Comment out/remove adjustwidth environment if table fits in text column.
\centering

\begin{tabular}{|c|c|c|c|c|c|c|c|c|c|c|c|}
\hline
\begin{tabular}[c]{@{}c@{}}Problem \\ name\end{tabular}                                                 & Difficulty & Laguage & \begin{tabular}[c]{@{}c@{}}Number \\ of \\ solutions \\ that \\ worked\end{tabular} & \begin{tabular}[c]{@{}c@{}}If failed-\\ number \\ of test \\ cases\end{tabular} & \begin{tabular}[c]{@{}c@{}}Runtime \\ (ms)\end{tabular} & \begin{tabular}[c]{@{}c@{}}Memory\\  (MB)\end{tabular} & \begin{tabular}[c]{@{}c@{}}Performance \\ (beats) \\ (\%)\end{tabular} & \begin{tabular}[c]{@{}c@{}}Memory \\ (beats) \\ (\%)\end{tabular} & \begin{tabular}[c]{@{}c@{}}Time \\ Complexity\end{tabular}                         & \begin{tabular}[c]{@{}c@{}}Space \\ Complexity\end{tabular} & \begin{tabular}[c]{@{}c@{}}ID \\ LeetCode\\ submission\end{tabular} \\ \hline
\begin{tabular}[c]{@{}c@{}}2656. \\ Maximum \\ Sum With \\ Exactly \\ K Elements\end{tabular}           & Easy       & C++     & 1/1                                                                                 & -                                                                               & 47                                                      &                                   70.6                     & 70.22                                                                  & 68.69                                                             & \begin{tabular}[c]{@{}c@{}}O(n log n + \\ k\textasciicircum{}2 log n)\end{tabular} & O(n)                                                        & 951082025                                                           \\ \hline
\begin{tabular}[c]{@{}c@{}}2656. \\ Maximum \\ Sum \\ With Exactly \\ K Elements\end{tabular}           & Easy       & C++     & 2/3                                                                                 & -                                                                               & 62                                                      & 71.7                                                   & 21.75                                                                  & 12.19                                                             & \begin{tabular}[c]{@{}c@{}}O(n + \\ k * log n)\end{tabular}                        & O(n)                                                        & 946957357                                                           \\ \hline
\begin{tabular}[c]{@{}c@{}}2657. \\ Find the \\ Prefix \\ Common \\ Array of \\ Two Arrays\end{tabular} & Medium     & C++     & 1/1                                                                                 & -                                                                               & 53                                                      & 80.9                                                   & 78.35                                                                  & 97.47                                                             & O(n\textasciicircum{}3)                                                            & O(n)                                                        & 951078174                                                           \\ \hline
\begin{tabular}[c]{@{}c@{}}2658. \\ Maximum \\ Number \\ of Fish \\ in a Grid\end{tabular}              & Medium     & C++     & 1/1                                                                                 & -                                                                               & 84                                                      & 88.4                                                   & 69.14                                                                  & 88.14                                                             & O((mn)²)                                                                           & O(mn)                                                       & 951076414                                                           \\ \hline
\begin{tabular}[c]{@{}c@{}}2659. \\ Make \\ Array \\ Empty\end{tabular}                                 & Hard       & C++     & 0                                                                                   & 505/514                                                                         & -                                                       & -                                                      & -                                                                      & -                                                                 & O(n\textasciicircum{}2)                                                            & O(n)                                                        & 951086982                                                           \\ \hline
\end{tabular} %
}
\caption{Result analysis of coding solutions provided by ChatGPT.}
\end{table*}

For some problems, ChatGPT-4 successfully generated a working solution on the first attempt. However, for others, multiple iterations were necessary to generate a solution that could pass all test cases. It is noteworthy that several software developers also had to submit multiple solutions before producing one that was accepted by Leetcode, passing all test cases.

The most challenging problem proved resistant, with none of the over 50 solutions generated (in various programming languages) able to pass all test cases. In this instance, the most effective solution generated by ChatGPT managed to pass 505 out of 514 test cases, falling short due to time limit exceedances for larger inputs. We posit that the subtlety of this question lies in the paradox between the textual description's suggestion of an algorithm to remove elements from an array and a hint indicating that the solution does not necessarily require the deletion or movement of elements, focusing instead on the array length. Essentially, it is anticipated that the developer's solution deviates from the direct interpretation of the problem description. Sobania et al. \cite{sobania2022choose}, in their evaluation of ChatGPT's code-generation capabilities against other similar solutions, noted that several problems which automated solutions could not solve were characterized by problem descriptions that were too ambiguous to be definitively solved, even by human programmers.

\subsection{Experiment - Results}

As detailed in Table \ref{tab:chatgptresults}, the solutions generated by ChatGPT-4 surpass 70.22\%, 78.35\%, and 69.14\% of existing solutions for each respective problem. In terms of memory efficiency, the solutions given by ChatGPT-4 outpace 68.69\%, 97.47\%, and 88.14\% of others, respectively. However, this comparison encompasses all previously provided solutions, thus lacking clarity on how these automated solutions perform relative to software engineers with varying expertise levels. 

We ran the solutions supplied by both groups of developers to ascertain the specific runtime and memory-efficiency of their code. Table 2 showcases a selection of the results derived from the solutions that developers produced for the least complex problem. Tables 3, 4, and 5 address this by exclusively considering the solutions provided by our selected participants to the easy, first medium, and second medium problems, respectively \footnote{The code provided by each participant and the submission links are available at:\\
\href{https://docs.google.com/spreadsheets/d/1Fqlz5kWQ7-Einxy5FRoLEeOXw_fBXulbO1qKbcorvyI/edit?usp=sharing}{Google Drive form (accessed on May 15, 2023)}}.

\label{tab:participant-easy}
\begin{table*}[!htb]

\scalebox{0.90}{% Comment out/remove adjustwidth environment if table fits in text column.
\centering

\begin{tabular}{|l|l|l|l|l|l|l|l|l|l|}
\hline
\textbf{Group} & \textbf{\begin{tabular}[c]{@{}l@{}}Developer\\ Num\end{tabular}} & \textbf{\begin{tabular}[c]{@{}l@{}}Global \\ Ranking\end{tabular}} & \textbf{\begin{tabular}[c]{@{}l@{}}Num of \\ C++ \\ problems \\ solved\end{tabular}} & \textbf{\begin{tabular}[c]{@{}l@{}}Contest \\ Position\end{tabular}} & \textbf{\begin{tabular}[c]{@{}l@{}}Top \\ (\%)\end{tabular}} & \textbf{\begin{tabular}[c]{@{}l@{}}Runtime \\ (ms)\end{tabular}} & \textbf{\begin{tabular}[c]{@{}l@{}}Memory \\ (MB)\end{tabular}} & \textbf{\begin{tabular}[c]{@{}l@{}}Performance\\ (beats) \\ (\%)\end{tabular}} & \textbf{\begin{tabular}[c]{@{}l@{}}Memory \\ (beats) \\ (\%)\end{tabular}} \\ \hline
A              & 5                                                                & 21                                                                 & 2426                                                                                 & 7                                                                    & 0.01                                                         & 29                                                               & 70.7                                                            & 98.27                                                                          & 29.83                                                                      \\ \hline
A              & 8                                                                & 1335                                                               & 55                                                                                   & 18                                                                   & 0.37                                                         & 36                                                               & 70.7                                                            & 93.71                                                                          & 68.73                                                                      \\ \hline
A              & 9                                                                & 2,915                                                              & 1098                                                                                 & 693                                                                  & 1.17                                                         & 59                                                               & 70.7                                                            & 30.22                                                                          & 29.83                                                                      \\ \hline
B              & 17                                                               & 334,617                                                            & 67                                                                                   & 12477                                                                & 85.17                                                        & 70                                                               & 70.6                                                            & 9.8                                                                            & 68.73                                                                      \\ \hline
B              & 18                                                               & 351,789                                                            & 140                                                                                  & 7117                                                                 & 89.45                                                        & 62                                                               & 70.8                                                            & 21.77                                                                          & 13.47                                                                      \\ \hline
B              & 19                                                               & 354,535                                                            & 82                                                                                   & 12487                                                                & 90.14                                                        & 52                                                               & 70.7                                                            & 52.49                                                                          & 29.83                                                                      \\ \hline
\end{tabular}%
}
\caption{A selection of solution  instances obtained from responses to the easy problem by the chosen participants (Groups A = Experienced; B = Novice).}
\end{table*}

\begin{table*}[!htb]

\scalebox{0.76}{% Comment out/remove adjustwidth environment if table fits in text column.
\centering
\begin{tabular}{|c|c|cc|c|ccc|c|ccc}
\hline
\textbf{Variable}                                                            & \textbf{\begin{tabular}[c]{@{}c@{}}N \\ samples\end{tabular}} & \multicolumn{1}{c|}{\textbf{\begin{tabular}[c]{@{}c@{}}Degrees \\ of \\ freedom\\ (n-1)\end{tabular}}} & \textbf{\begin{tabular}[c]{@{}c@{}}t critical\\ value\\ (.99\%)\end{tabular}} & \textbf{\begin{tabular}[c]{@{}c@{}}Runtime\\ -\\ Best Value\end{tabular}} & \multicolumn{1}{c|}{\textbf{\begin{tabular}[c]{@{}c@{}}Runtime\\ -\\ Mean\end{tabular}}} & \multicolumn{1}{c|}{\textbf{\begin{tabular}[c]{@{}c@{}}Runtime\\ -\\ Standard\\ deviation\end{tabular}}} & \textbf{\begin{tabular}[c]{@{}c@{}}Runtime\\ -\\ tstatistic\end{tabular}} & \textbf{\begin{tabular}[c]{@{}c@{}}Memory\\ -\\ Best \\ Value\end{tabular}} & \multicolumn{1}{c|}{\textbf{\begin{tabular}[c]{@{}c@{}}Memory\\ -\\ Mean\end{tabular}}} & \multicolumn{1}{c|}{\textbf{\begin{tabular}[c]{@{}c@{}}Memory\\ -\\ Std\end{tabular}}} & \multicolumn{1}{c|}{\textbf{\begin{tabular}[c]{@{}c@{}}Memory\\ -\\ tstatistic\end{tabular}}} \\ \hline
Programmers                                                                  & 38                                                            & \multicolumn{1}{c|}{37}                                                                                & -2.43                                                                         & 29                                                                        & \multicolumn{1}{c|}{54.63}                                                               & \multicolumn{1}{c|}{16.60}                                                                               & -2.83                                                                     & 70.6                                                                        & \multicolumn{1}{c|}{71.45}                                                              & \multicolumn{1}{c|}{2.84}                                                              & \multicolumn{1}{c|}{-1.86}                                                                    \\ \hline
\begin{tabular}[c]{@{}c@{}}Experienced\\ contest \\ programmers\end{tabular} & 20                                                            & \multicolumn{1}{c|}{19}                                                                                & -2.53                                                                         & 29                                                                        & \multicolumn{1}{c|}{55}                                                                  & \multicolumn{1}{c|}{21.70}                                                                               & -1.64                                                                     & 70.6                                                                        & \multicolumn{1}{c|}{72.035}                                                             & \multicolumn{1}{c|}{3.83}                                                              & \multicolumn{1}{c|}{-1.67}                                                                    \\ \hline
\begin{tabular}[c]{@{}c@{}}Novice\\ contest\\ programmers\end{tabular}       & 18                                                            & \multicolumn{1}{c|}{17}                                                                                & -2.56                                                                         & 35                                                                        & \multicolumn{1}{c|}{54.22}                                                               & \multicolumn{1}{c|}{8.57}                                                                                & -3.57                                                                     & 70.6                                                                        & \multicolumn{1}{c|}{70.81}                                                              & \multicolumn{1}{c|}{0.55}                                                              & \multicolumn{1}{c|}{-1.66}                                                                    \\ \hline
ChatGPT-4                                                                    & 1                                                             & \multicolumn{2}{c|}{}                                                                                                                                                                  & 47                                                                        & \multicolumn{3}{c|}{}                                                                                                                                                                                                                                                           & 70.6                                                                        & \multicolumn{3}{c}{}                                                                                                                                                                                                                                                             \\ \cline{1-2} \cline{5-5} \cline{9-9}
\end{tabular}%
}
\caption{Easy problem (2656. Maximum Sum With Exactly K Elements): data to perform test statistic - performance and memory efficiency.}
\end{table*} \label{tab:easyresult}

\begin{table*}[!htb]

\scalebox{0.76}{% Comment out/remove adjustwidth environment if table fits in text column.
\centering
\begin{tabular}{|c|c|cc|c|ccc|c|ccc}
\hline
\textbf{Variable}                                                            & \textbf{\begin{tabular}[c]{@{}c@{}}N \\ samples\end{tabular}} & \multicolumn{1}{c|}{\textbf{\begin{tabular}[c]{@{}c@{}}Degrees \\ of \\ freedom\\ (n-1)\end{tabular}}} & \textbf{\begin{tabular}[c]{@{}c@{}}t critical\\ value\\ (.99\%)\end{tabular}} & \textbf{\begin{tabular}[c]{@{}c@{}}Runtime\\ -\\ Best Value\end{tabular}} & \multicolumn{1}{c|}{\textbf{\begin{tabular}[c]{@{}c@{}}Runtime\\ -\\ Mean\end{tabular}}} & \multicolumn{1}{c|}{\textbf{\begin{tabular}[c]{@{}c@{}}Runtime\\ -\\ Standard\\ deviation\end{tabular}}} & \textbf{\begin{tabular}[c]{@{}c@{}}Runtime\\ -\\ tstatistic\end{tabular}} & \textbf{\begin{tabular}[c]{@{}c@{}}Memory\\ -\\ Best \\ Value\end{tabular}} & \multicolumn{1}{c|}{\textbf{\begin{tabular}[c]{@{}c@{}}Memory\\ -\\ Mean\end{tabular}}} & \multicolumn{1}{c|}{\textbf{\begin{tabular}[c]{@{}c@{}}Memory\\ -\\ Std\end{tabular}}} & \multicolumn{1}{c|}{\textbf{\begin{tabular}[c]{@{}c@{}}Memory\\ -\\ tstatistic\end{tabular}}} \\ \hline
Programmers                                                                  & 34                                                            & \multicolumn{1}{c|}{33}                                                                                & -2.44                                                                         & 48                                                                        & \multicolumn{1}{c|}{118.5}                                                               & \multicolumn{1}{c|}{123.03}                                                                              & -3.10                                                                     & 80.7                                                                        & \multicolumn{1}{c|}{91.32}                                                              & \multicolumn{1}{c|}{24.09}                                                             & \multicolumn{1}{c|}{-2.52}                                                                    \\ \hline
\begin{tabular}[c]{@{}c@{}}Experienced\\ contest \\ programmers\end{tabular} & 19                                                            & \multicolumn{1}{c|}{18}                                                                                & -2.55                                                                         & 48                                                                        & \multicolumn{1}{c|}{83.84}                                                               & \multicolumn{1}{c|}{67.03}                                                                               & -2.00                                                                     & 80.7                                                                        & \multicolumn{1}{c|}{87.66}                                                              & \multicolumn{1}{c|}{15.86}                                                             & \multicolumn{1}{c|}{-1.85}                                                                    \\ \hline
\begin{tabular}[c]{@{}c@{}}Novice\\ contest\\ programmers\end{tabular}       & 15                                                            & \multicolumn{1}{c|}{14}                                                                                & -2.62                                                                         & 56                                                                        & \multicolumn{1}{c|}{162.4}                                                               & \multicolumn{1}{c|}{161.89}                                                                              & -2.61                                                                     & 81.1                                                                        & \multicolumn{1}{c|}{95.95}                                                              & \multicolumn{1}{c|}{31.68}                                                             & \multicolumn{1}{c|}{-1.84}                                                                    \\ \hline
ChatGPT-4                                                                    & 1                                                             & \multicolumn{2}{c|}{}                                                                                                                                                                  & 53                                                                        & \multicolumn{3}{c|}{}                                                                                                                                                                                                                                                           & 80.9                                                                        & \multicolumn{3}{c}{}                                                                                                                                                                                                                                                             \\ \cline{1-2} \cline{5-5} \cline{9-9}
\end{tabular}%
}
\caption{Medium problem (2657. Find the Prefix Common Array of Two Arrays): data to perform test statistic - performance and memory efficiency.}
\end{table*} \label{tab:medium1result}

\begin{table*}[!htb]

\scalebox{0.76}{% Comment out/remove adjustwidth environment if table fits in text column.
\centering
\begin{tabular}{|c|c|cc|c|ccc|c|ccc}
\hline
\textbf{Variable}                                                            & \textbf{\begin{tabular}[c]{@{}c@{}}N \\ samples\end{tabular}} & \multicolumn{1}{c|}{\textbf{\begin{tabular}[c]{@{}c@{}}Degrees \\ of \\ freedom\\ (n-1)\end{tabular}}} & \textbf{\begin{tabular}[c]{@{}c@{}}t critical\\ value\\ (.99\%)\end{tabular}} & \textbf{\begin{tabular}[c]{@{}c@{}}Runtime\\ -\\ Best Value\end{tabular}} & \multicolumn{1}{c|}{\textbf{\begin{tabular}[c]{@{}c@{}}Runtime\\ -\\ Mean\end{tabular}}} & \multicolumn{1}{c|}{\textbf{\begin{tabular}[c]{@{}c@{}}Runtime\\ -\\ Standard\\ deviation\end{tabular}}} & \textbf{\begin{tabular}[c]{@{}c@{}}Runtime\\ -\\ tstatistic\end{tabular}} & \textbf{\begin{tabular}[c]{@{}c@{}}Memory\\ -\\ Best \\ Value\end{tabular}} & \multicolumn{1}{c|}{\textbf{\begin{tabular}[c]{@{}c@{}}Memory\\ -\\ Mean\end{tabular}}} & \multicolumn{1}{c|}{\textbf{\begin{tabular}[c]{@{}c@{}}Memory\\ -\\ Std\end{tabular}}} & \multicolumn{1}{c|}{\textbf{\begin{tabular}[c]{@{}c@{}}Memory\\ -\\ tstatistic\end{tabular}}} \\ \hline
Programmers                                                                  & 32                                                            & \multicolumn{1}{c|}{31}                                                                                & -2.45                                                                         & 56                                                                        & \multicolumn{1}{c|}{154.71}                                                              & \multicolumn{1}{c|}{95.34}                                                                               & -4.19                                                                     & 88.4                                                                        & \multicolumn{1}{c|}{110.64}                                                             & \multicolumn{1}{c|}{26.77}                                                             & \multicolumn{1}{c|}{-4.69}                                                                    \\ \hline
\begin{tabular}[c]{@{}c@{}}Experienced\\ contest \\ programmers\end{tabular} & 19                                                            & \multicolumn{1}{c|}{18}                                                                                & -2.55                                                                         & 56                                                                        & \multicolumn{1}{c|}{138.57}                                                              & \multicolumn{1}{c|}{102.46}                                                                              & -2.32                                                                     & 88.4                                                                        & \multicolumn{1}{c|}{105.32}                                                             & \multicolumn{1}{c|}{26.57}                                                             & \multicolumn{1}{c|}{-2.77}                                                                    \\ \hline
\begin{tabular}[c]{@{}c@{}}Novice\\ contest\\ programmers\end{tabular}       & 13                                                            & \multicolumn{1}{c|}{12}                                                                                & -2.68                                                                         & 76                                                                        & \multicolumn{1}{c|}{178.30}                                                              & \multicolumn{1}{c|}{81.96}                                                                               & -4.148                                                                    & 88.4                                                                        & \multicolumn{1}{c|}{118.40}                                                             & \multicolumn{1}{c|}{26.12}                                                             & \multicolumn{1}{c|}{-4.141}                                                                   \\ \hline
ChatGPT-4                                                                    & 1                                                             & \multicolumn{2}{c|}{}                                                                                                                                                                  & 84                                                                        & \multicolumn{3}{c|}{}                                                                                                                                                                                                                                                           & 88.4                                                                        & \multicolumn{3}{c}{}                                                                                                                                                                                                                                                             \\ \cline{1-2} \cline{5-5} \cline{9-9}
\end{tabular}%
}
\caption{Medium problem (2658. Maximum Number of Fish in a Grid): data to perform test statistic - performance and memory efficiency.}
\end{table*} \label{tab:medium2result}

Pertaining to the high-difficulty problem, it remained unsolved by both ChatGPT and novice contest programmers. Only the experienced contest programmers were successful in delivering an effective solution for this coding problem.

\subsubsection{Hypothesis Testing}
% VER a partir da pg 25. muito bom: https://www.sheffield.ac.uk/polopoly_fs/1.104345!/file/Scope_tutorial_manual.pdf Acho q posso aplicar Chi-squared test pg 58 - Chi square é aconselhavel para muitos dados

In this part, we investigate the hypotheses regarding the evaluation of solutions' performance and memory efficiency, as introduced in subsection \ref{sec:questionsexp1}. Consequently, we conducted statistical analyses, following the methods laid out by Peck and Devore \cite{peck2011statistics}, based on the metrics displayed in the three tables from Section \ref{tab:easyresult}.

We divided the experiment's outcomes into two categories: those from experienced contest programmers and those from novice contest programmers. Subsequently, we computed the mean and standard deviation of the results for each group. These statistical measures were then juxtaposed with the results obtained using ChatGPT. For instance, the first hypothesis (H - RQ1) posits that the solution generated by ChatGPT enhances application performance compared to solutions proposed by expert software engineers, leading to a reduction in runtime. The assertion, therefore, is that the mean runtime of expert software engineers' solutions surpasses that of the ChatGPT solution, which recorded 47ms for the easy problem, and 53ms and 84ms for the medium problems respectively.

Using a statistical significance level \cite{peck2011statistics} of 0.01 (the chance of one in 100 of making an error), we computed the test statistic (\begin{math}  t-statistic\end{math}) for each one of the difficulty levels, as follows \cite{peck2011statistics}:

\begin{equation}
t-statistic: t =\frac{(\overline{x} - hypothesized value)}{(\frac{\sigma}{\sqrt {n}} )}
\label{eq:tstatictis}
\end{equation}

Based on t-statistic theory, we can confidently reject our null hypothesis if the \begin{math}  t-statistic\end{math} value falls below the negative \begin{math} t - critical value \end{math} (threshold) \cite{peck2011statistics}. This implies that if we had assessed the entire sample of selected participants without categorizing them into two groups, we could have found evidence suggesting that automated solutions, like ChatGPT, could potentially outperform programmers in general. As demonstrated in the three result tables, the \begin{math}runtime  t-statistic\end{math} value is below the negative \begin{math} t - critical value \end{math} value in all cases when considering the entire group of programmers. The same holds true for the \begin{math}memory  t-statistic\end{math} in the medium-level problems.

\subsection{Discussion}
In our empirical investigation, where we evaluated the performance and memory efficiency of solutions offered by programmers of varying experience levels, we validated three alternative hypotheses while three were rejected:

\subsubsection*{	Accepted:}
\begin{enumerate}
\item	For easy and medium-level problems, ChatGPT enhances the performance of coding solutions in comparison to those provided by novice contest programmers.
\item	For medium-level problems, ChatGPT enhances memory efficiency in coding solutions in comparison to those provided by both experienced and novice contest programmers.
\end{enumerate}

\subsubsection*{	Rejected:}	
\begin{enumerate}
	\item ChatGPT enhances the performance of coding solutions for easy, medium, or hard-level problems when compared to those provided by experienced contest programmers.
 \item For easy and hard-level coding problems, ChatGPT demonstrates superior memory efficiency when compared to the solutions offered by either novice or experienced contest programmers.
\end{enumerate}

These findings suggest that automated solutions, such as ChatGPT, may outperform software engineers in certain software engineering tasks. Specifically, ChatGPT demonstrated superior performance over novice contest programmers in solving easy and medium-level problems and also exhibited better memory efficiency in one of the medium-level problems.

However, we found no evidence to assert that ChatGPT surpasses the performance of solutions provided by experienced contest programmers. This insight is notable given the significant research interest in the automation of software development tasks.

In summary, our study suggests a nuanced relationship between the performance of software engineers and AI-based solutions: in certain scenarios, software engineers excel, while in others, AI proves superior. This underscores the importance of understanding the unique strengths of both human and automated approaches, facilitating more effective collaborative work and task allocation processes \cite{nascimento2021approach}. The findings also emphasize the need for AI systems with adaptable degrees of automation, in line with the perspective offered by Melo et al. \cite{melo2022understanding}. Within a software engineering context, this suggests adjusting the level of AI automation based on both the experience of the developer and the quality requirements of the task at hand.

\subsection{Threats to Validity}

The exact training and testing data employed by ChatGPT remains undisclosed, meaning we cannot ascertain if our queries' precise solutions already exist within the data. Consequently, the specific coding challenges used to train their tool remain unidentified. Even though we opted for a recent LeetCode contest with novel problems, we cannot ensure that we did not test problems similar to or the same as those used in training their algorithm. As such, the tool may not be creating a fresh solution but could be retrieving a previously stored solution for a specific problem.

Moreover, there's the potential that software engineers have previously leveraged automated systems to submit questions to LeetCode, without giving due credit. As such, we cannot assert that the comparisons made were exclusively with solutions provided by developers. For instance, Golzadeh et al. have presented evidence suggesting that bots regularly feature among the most active contributors on GitHub, despite GitHub not recognizing their contributions \cite{golzadeh2022recognizing}. To mitigate this issue, we implemented participant selection.

\section{Conclusion and Future Work}

Several researchers have proposed the use of AI systems to automate software engineering tasks. However, most of these approaches do not direct efforts toward asking whether AI-based procedures have higher success rates than current standard and manual practices. A relevant question in this potential line of investigation is: ``Could a software engineer solve a specific development task better than an automated system?". Indeed, it is fundamental to evaluate which tasks are better performed by engineers or AI procedures so that they can work together more effectively and also provide more insight into novel human-in-the-loop AI approaches to support SE tasks.

Though there is conjecture that AI-based computation could enhance productivity and potentially replace software engineers in software development, current empirical evidence supporting this claim is scant. Indeed, a limited number of papers offer empirical investigations into the application of machine-learning techniques in Software Engineering. This paper introduces an empirical study examining the utilization of automated strategies like ChatGPT to automate a task in Software Engineering, specifically, solving coding problems. Moreover, as inherent in experimental studies, even with careful design and execution, certain factors could pose threats to experimental validity. One such potential threat includes the precise training and testing data employed by ChatGPT.

Our empirical study uncovered that automated systems like ChatGPT can, in certain instances, surpass the performance of novice software engineers in specific tasks. This superiority was particularly evident in the solving of easy and medium-level problems, where ChatGPT's performance consistently exceeded that of novice contest programmers. Moreover, the AI-based solution demonstrated improved memory efficiency for a medium-level problem. In contrast, we found no substantial evidence to suggest that ChatGPT could outdo experienced contest programmers in terms of solution performance. In essence, our study reveals a dynamic interplay between human and AI performance in software engineering tasks, highlighting the need for different task allocation processes. This encourages a collaborative approach, fine-tuning AI assistance based on developer expertise and task quality requirements.

This empirical investigation ought to explore the potential for automation in software engineering tasks extending beyond the realm of problem-solving in coding. Future work to extend the proposed experiment includes: (i) conducting further empirical studies to assess other SE tasks, such as design, maintenance, testing, and project management; (ii) experimenting with other AI approaches, such as unsupervised machine-learning algorithms; and (iii) using different criteria to evaluate task execution, addressing different qualitative or quantitative methodologies. Possible tasks that could be investigated (refer to (i)) include testing tasks (e.g. comparing the number, type and difficulty level of faults that were identified by developers), designing tasks (i.e. accessing system usability), maintenance tasks (i.e. accessing continuous performance), and project management tasks (i.e. evaluating the level of satisfaction of developers in the task allocation process). 

\section*{Acknowledgment}
 This work was supported by the Natural Sciences and Engineering Research Council of Canada (NSERC), and the Centre for Community Mapping (COMAP).


%\bibliographystyle{ACM-Reference-Format}
%\bibliographystyle{unsrtnat}
\bibliographystyle{unsrt}
\bibliography{ieee-transaction,new-refs,bib-model-human-loop.bib}

\begin{thebibliography}{10}

\bibitem{openai2023gpt4}
OpenAI.
\newblock Gpt-4 technical report.
\newblock {\em https://doi.org/10.48550/arXiv.2303.08774}, 2023.

\bibitem{eloundou2023gpts}
Tyna Eloundou, Sam Manning, Pamela Mishkin, and Daniel Rock.
\newblock Gpts are gpts: An early look at the labor market impact potential of
  large language models, 2023.

\bibitem{white2023chatgpt}
Jules White, Sam Hays, Quchen Fu, Jesse Spencer-Smith, and Douglas~C Schmidt.
\newblock Chatgpt prompt patterns for improving code quality, refactoring,
  requirements elicitation, and software design.
\newblock {\em https://doi.org/10.48550/arXiv.2303.07839}, 2023.

\bibitem{sobania2023analysis}
Dominik Sobania, Martin Briesch, Carol Hanna, and Justyna Petke.
\newblock An analysis of the automatic bug fixing performance of chatgpt.
\newblock {\em https://doi.org/10.48550/ARXIV.2301.08653}, 2023.

\bibitem{sarro2023automated}
Federica Sarro.
\newblock Automated optimisation of modern software system properties.
\newblock In {\em Proceedings of the 2023 ACM/SPEC International Conference on
  Performance Engineering}, pages 3--4, 2023.

\bibitem{melo2023supporting}
Glaucia Melo, Luis~Fernando Lins, Paulo Alencar, and Donald Cowan.
\newblock Supporting contextual conversational agent-based software
  development.
\newblock {\em International Conference on Software Engineering}, 2023.

\bibitem{imai2022github}
Saki Imai.
\newblock Is github copilot a substitute for human pair-programming? an
  empirical study.
\newblock In {\em Proceedings of the ACM/IEEE 44th International Conference on
  Software Engineering: Companion Proceedings}, pages 319--321, 2022.

\bibitem{golzadeh2022recognizing}
Mehdi Golzadeh, Tom Mens, Alexandre Decan, Eleni Constantinou, and Natarajan
  Chidambaram.
\newblock Recognizing bot activity in collaborative software development.
\newblock {\em IEEE Software}, 39(5):56--61, 2022.

\bibitem{bender2021dangers}
Emily~M Bender, Timnit Gebru, Angelina McMillan-Major, and Shmargaret
  Shmitchell.
\newblock On the dangers of stochastic parrots: Can language models be too big?
\newblock In {\em Proceedings of the 2021 ACM conference on fairness,
  accountability, and transparency}, pages 610--623, 2021.

\bibitem{georgiou2022green}
Stefanos Georgiou, Maria Kechagia, Tushar Sharma, Federica Sarro, and Ying Zou.
\newblock Green ai: Do deep learning frameworks have different costs?
\newblock In {\em Proceedings of the 44th International Conference on Software
  Engineering}, pages 1082--1094, 2022.

\bibitem{pearce2022asleep}
Hammond Pearce, Baleegh Ahmad, Benjamin Tan, Brendan Dolan-Gavitt, and Ramesh
  Karri.
\newblock Asleep at the keyboard? assessing the security of github copilot’s
  code contributions.
\newblock In {\em 2022 IEEE Symposium on Security and Privacy (SP)}, pages
  754--768. IEEE, 2022.

\bibitem{10.1145/3457607}
Ninareh Mehrabi, Fred Morstatter, Nripsuta Saxena, Kristina Lerman, and Aram
  Galstyan.
\newblock A survey on bias and fairness in machine learning.
\newblock {\em ACM Comput. Surv.}, 54(6), jul 2021.

\bibitem{chen2021evaluating}
Mark Chen, Jerry Tworek, Heewoo Jun, Qiming Yuan, Henrique Ponde de~Oliveira
  Pinto, Jared Kaplan, Harri Edwards, Yuri Burda, Nicholas Joseph, Greg
  Brockman, et~al.
\newblock Evaluating large language models trained on code.
\newblock {\em arXiv preprint arXiv:2107.03374}, 2021.

\bibitem{inala2022fault}
Jeevana~Priya Inala, Chenglong Wang, Mei Yang, Andres Codas, Mark
  Encarnaci{\'o}n, Shuvendu Lahiri, Madanlal Musuvathi, and Jianfeng Gao.
\newblock Fault-aware neural code rankers.
\newblock {\em Advances in Neural Information Processing Systems},
  35:13419--13432, 2022.

\bibitem{nascimento2018toward}
Nathalia Nascimento, Paulo Alencar, Carlos Lucena, and Donald Cowan.
\newblock Toward human-in-the-loop collaboration between software engineers and
  machine learning algorithms.
\newblock In {\em 2018 IEEE International Conference on Big Data (Big Data)},
  pages 3534--3540. IEEE, 2018.

\bibitem{easterbrook2008selecting}
Steve Easterbrook, Janice Singer, Margaret-Anne Storey, and Daniela Damian.
\newblock Selecting empirical methods for software engineering research.
\newblock {\em Guide to advanced empirical software engineering}, pages
  285--311, 2008.

\bibitem{nguyen2022empirical}
Nhan Nguyen and Sarah Nadi.
\newblock An empirical evaluation of github copilot's code suggestions.
\newblock In {\em Proceedings of the 19th International Conference on Mining
  Software Repositories}, pages 1--5, 2022.

\bibitem{li2022competition}
Yujia Li, David Choi, Junyoung Chung, Nate Kushman, Julian Schrittwieser,
  R{\'e}mi Leblond, Tom Eccles, James Keeling, Felix Gimeno, Agustin Dal~Lago,
  et~al.
\newblock Competition-level code generation with alphacode.
\newblock {\em Science}, 378(6624):1092--1097, 2022.

\bibitem{lertbanjongngam2022empirical}
Sila Lertbanjongngam, Bodin Chinthanet, Takashi Ishio, Raula~Gaikovina Kula,
  Pattara Leelaprute, Bundit Manaskasemsak, Arnon Rungsawang, and Kenichi
  Matsumoto.
\newblock An empirical evaluation of competitive programming ai: A case study
  of alphacode.
\newblock In {\em 2022 IEEE 16th International Workshop on Software Clones
  (IWSC)}, pages 10--15. IEEE, 2022.

\bibitem{brooks1987no}
F~Brooks and HJ~Kugler.
\newblock {\em No silver bullet}.
\newblock April, 1987.

\bibitem{sjoberg2008building}
Dag~IK Sj{\o}berg, Tore Dyb{\aa}, Bente~CD Anda, and Jo~E Hannay.
\newblock Building theories in software engineering.
\newblock {\em Guide to advanced empirical software engineering}, pages
  312--336, 2008.

\bibitem{sobania2022choose}
Dominik Sobania, Martin Briesch, and Franz Rothlauf.
\newblock Choose your programming copilot: a comparison of the program
  synthesis performance of github copilot and genetic programming.
\newblock In {\em Proceedings of the Genetic and Evolutionary Computation
  Conference}, pages 1019--1027, 2022.

\bibitem{peck2011statistics}
Roxy Peck and Jay Devore.
\newblock {\em Statistics: The Exploration \& Analysis of Data}.
\newblock Nelson Education, 2011.

\bibitem{nascimento2021approach}
Nathalia Nascimento, Paulo Alencar, and Donald Cowan.
\newblock An approach to support human-in-the-loop big data software
  development projects.
\newblock In {\em 2021 IEEE International Conference on Big Data (Big Data)},
  pages 2319--2326. IEEE, 2021.

\bibitem{melo2022understanding}
Glaucia Melo, Nathalia Nascimento, Paulo Alencar, and Donald Cowan.
\newblock Understanding levels of automation in human-machine collaboration.
\newblock In {\em 2022 IEEE International Conference on Big Data (Big Data)},
  pages 3952--3958. IEEE, 2022.

\end{thebibliography}

\end{document}